# Autonomous Small-Angle Scattering for Accelerated Soft Material Formulation Optimization

Tyler B. Martin[*], Duncan Sutherland, Austin McDannald, A. Gilad Kusne, Peter A. Beaucage*

1. National Institute of Standards and Technology, Gaithersburg MD, 20899

* corresponding authors: tyler.martin@nist.gov and peter.beaucage@nist.gov

# Abstract

The pace of soft material formulation (re)development and design is rapidly increasing as both consumers and new legislation demand products that do less harm to the environment while maintaining high standards of performance. To meet this need, we have developed the Autonomous Formulation Lab (AFL), a platform that can automatically prepare and measure the microstructure of liquid formulations using small-angle neutron and X-ray scattering and, soon, a variety of other techniques. Here, we describe the design, philosophy, tuning, and validation of our active learning agent that guides the course of AFL experiments. We show how our extensive *in silico* tuning results in an efficient agent that is robust to both the number of measurements and signal to noise variation. Finally, we experimentally validate our virtually tuned agent by addressing a model formulation problem: replacing a petroleum-derived component with a natural analog. We show that the agent efficiently maps both formulations and how post hoc analysis of the measured data reveals the opportunity for further specialization of the agent. With the tuned and proven active learning agent, our autonomously guided AFL platform will accelerate the pace of discovery of liquid formulations and help speed us towards a greener future.

# I. Introduction

The timescale of innovation driven by fundamental understanding in materials development is often on the order of years. For example, the key technologies for carbon-fiber reinforced polymers (CRFP) were developed in the 1950s and 1960s, yet the first passenger airplanes with significant use of CFRP only found commercial application in the 2010s – a 50-year lag. Accelerating the materials innovation lifecycle from fundamental understanding to product-on-market is crucial to meeting the challenges of energy, water, health, and environment required by the growing global population. Efforts to accelerate this lifecycle, such as the Materials Genome Initiative,[1] have largely focused on harnessing advanced modeling and modern informatics to supplement human expertise in deciding which experiments to perform to produce the maximum measurement value from the minimum number of experiments. Such workflows and related *in silico* informatics approaches have been transformative in exploring the equilibrium phase diagrams of hard materials, such as metals and ceramics, and some non-equilibrium materials processing.

In contrast to metals and ceramics, where compositions with 5 elemental components are considered highly complex,[2, 3] liquid formulations regularly consist of tens to hundreds of components with the physics of structure formation driven by each component. Formulated products span the breadth of modern society, from pharmaceuticals, personal care products, and food and beverage to drilling fluids and industrial materials. Despite the success of advanced modeling and closed-loop studies in other areas, most formulations are developed using human expert intuition or at best, design-of-experiments (DOE) strategies focused on phenomenological or structural properties. This gap originates in several fundamental features of formulations: they are nearly always highly multicomponent, their structures are often far-from-equilibrium with significant processing pathway dependence, and their structure-property relationships are frequently highly complex and poorly theoretically understood. Take, as an example, a typical hair shampoo. At its core, the product is a simple ternary mixture of surfactant, oil (such as conditioners, fragrances, etc), and water. However, the real formulation almost always includes a blend of several ionic and non-ionic surfactants (*e.g.,* multiple sodium laureth sulfates (SLES) with different ethylene oxide unit lengths or cocamidopropyl betaine (CAPB)), the oil component includes tens to hundreds of different species with diverse chemical functionality and physics, and the water component includes salts and pH adjusters.  These components can interact to produce a variety of different microstructures from spherical and wormlike micelles to vesicles, which in turn produce a given final property, such as viscosity based on micelle entanglement or other macromolecular factors.  In such a complex landscape, small changes in single components (for example, changing which fragrance blend is used or the overall fragrance loading) can result in unexpected changes in final properties. It is extraordinarily difficult to know in advance how far a given formulation is from a performance boundary and the potential impact from a minor variation in a manufacturing parameter.  Fundamental structural understanding can be transformative to this process, but we generally lack physical theories both accurate enough and able to handle these complex formulation systems.

To address this complexity, we have developed a flexible and open automation platform, the Autonomous Formulation Laboratory (AFL), which is capable of automated preparation and measurement of liquid formulations via pipetting. It can be coupled to small-angle X-ray or neutron scattering instruments for structural 'ground truth' measurements together with benchtop performance data such as UV-vis, turbidity, and capillary viscometry.[4] While this platform provides enhanced reproducibility and throughput for measurements, the realization of its full potential for accelerated materials discovery requires data approaches that can, with minimal human input or training data, accurately ***interpret/label*** measurements, ***extrapolate*** those measurements into a statistically-derived phase diagram, and ***choose*** which next measurement to perform toward a specific scientific objective, such as determination of the desired phase boundaries, the discovery of the overall phase behavior and boundaries of an unknown system, or the optimization of a property of interest. The difference between an automated platform and an autonomous one is the decision-making *agent* that optimally guides the course of an experiment.

For hard materials science and small molecule chemistry, agent-guided closed-loop autonomous experiments have shown great value in discovering new materials, optimizing material properties, and mapping phase boundaries.[5-12] Many of these studies have moved past the simple application of black box machine learning techniques and have attempted to incorporate material or measurement physics into their agents. This can include incorporating thermodynamic constraints (*e.g.,* Gibbs phase rule), knowledge of crystallographic concepts like peak shifting, or knowledge about the potential mathematical descriptions of physical phenomena.[5, 8, 12] Others have sought to add explainability to their agents, such as through interpretable constraints applied to latent spaces of known variables.[13] Frameworks have been developed for allowing multiple autonomous agents to interactively collaborate.[14] Additionally, there is a growing conversation in the autonomous experimentation community on how to implement "human-machine teaming" concepts where autonomous agents collaborate with humans to combine the speed of autonomous agent decision making with human knowledge and intuition.[15-18]

Comparatively, the application of active learning and autonomous techniques to polymer and soft material systems is less developed. Several groups have identified reversible addition-fragmentation chain transfer (RAFT) polymer synthesis as being highly amenable to flow geometries and therefore automation.[19-22] These studies sought to optimize material properties by tuning the polymer synthesis to control the polymer sequence or molecular weight distributions of their polymers. There have been a smaller number of studies on polymer property optimization of polymer formulations.[23-27] These studies optimize material properties by tuning composition rather than synthesis. Of particular interest are studies that take industrial interests into account, such as the cost of the formulation chemistry.[26]

Here we report the development of a modular active learning agent that leverages small-angle scattering (SAS) measurements for phase discovery. Our agent is designed to be general-purpose such that it can be applied to scattering (and non-scattering) instruments and a range of material systems without prior knowledge of a system's phase behavior. In the following sections we discuss the design of the agent, our *in silico* tuning approach, and finally closed-loop experimental validation delivering a performance increase of as much as 25x compared to naïve grid searches.

## II. Agent Architecture

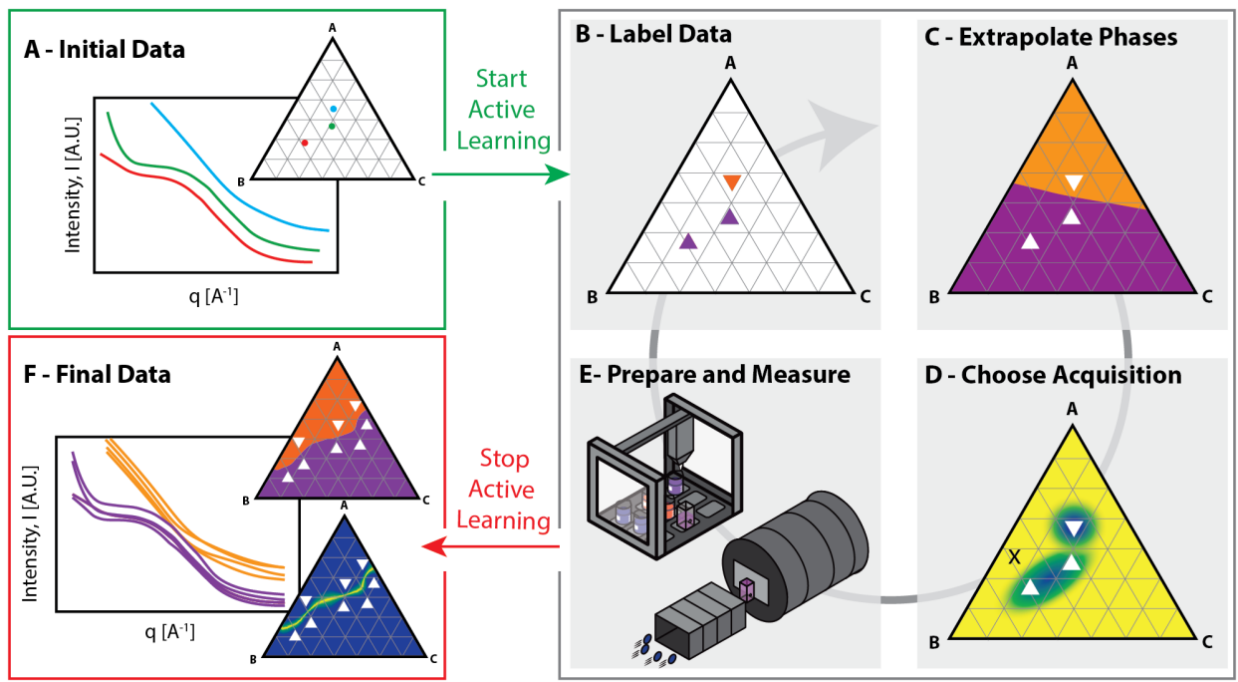

*Figure 1: Schematic overview of the active learning agent. (A) A phase mapping problem is posed to the agent in the form of a set of least two small-angle scattering data labeled with the composition of the sample. These data can originate from expertly chosen, literature gathered, or randomly-measured data. (B) These points are labeled and grouped using similarity and clustering analyses. (C) The labeled data is then extrapolated over the entire composition space using a Gaussian process classifier. (D) The next sample that best achieves the measurement goal is chosen using an acquisition function. (E) The chosen sample composition is robotically prepared and measured. (F) The loop is broken when the measurement goal is achieved with sufficient confidence from the AFL operator.*

### A. Label Data

Figure 1 is a schematic depiction of our agent's active learning process with the full details found in Section S1 of the Supplementary Information. Here we provide an overview of the agent's design.

The agent begins each iteration of the active learning loop by analyzing the current scattering data and assigning each pattern to a discrete phase group. Even for highly trained *human* practitioners, interpreting SAS data is a highly non-trivial task. Due to the inherent heterogeneity of soft materials and the nature of SAS measurements, many microstructures can produce the same or similar scattering patterns. Choosing from the plethora of available geometric, thermodynamic, and empirical models requires specialized knowledge of the system under study, and often the results of other non-SAS measurements. Even with an appropriately chosen model, fitting SAS models can be challenging and perfect fits to experimental data are rare due to non-idealities in the instrument and sample.

The most direct approach to automating SAS data analysis would be to develop a classifier that identifies one or more appropriate analytical models, fits those models, and then uses the fit parameters to identify phases. While ML classification models for SAS exist in the literature, they have mostly been applied to theoretical data and tuning these models for our applications would be challenging for all the reasons highlighted above.[28, 29] To overcome this challenge, we leverage a SAS-model-free approach that combines similarity analysis with a clustering routine that gathers the data into groups of similar scattering. This group identity of each measurement acts as the phase label for that composition, *i.e.* all data in cluster 1 belong to "phase 1". While this similarity approach to phase identification does not give the true structural label of the phase (*e.g.,* spherical micelles or lamellae) it does not require *a priori* knowledge of the phase behavior of a system before beginning a campaign.

The core of this approach lies in a mathematical similarity kernel. Tuning both the form and parameters of this kernel is necessary to tailor the autonomous agent to SAS measurements. Identifying a common similarity kernel across material systems is analogous to coming up with a generalized taxonomy for SAS data: in some contexts, scattering data should be sorted into "crystalline" and "amorphous" while other datasets might demand differentiation between amorphous states (cylindrical particles vs spherical, etc.). To identify candidates for generally applicable similarity kernels, we developed a virtual instrument platform that allowed us to screen >250 000 labeling pipelines and identify the best candidates for application. As described in Section III, we find several pipelines that provide accurate labeling but are also robust to various measurement details (*e.g.,* number of datapoints, measurement noise).

This similarity approach represents a highly general and adaptable method for analyzing and labeling SAS data. While we have focused on SAS, nothing in the above method is strongly specialized for this class of measurement. It is straightforward to test similarity metrics and clustering accuracy on pre-measured data, so adapting the featurization/labeling steps to a new measurement (e.g., spectroscopy or rheology) is not challenging. In addition, the similarity approach leaves an easy route towards incorporating multiple measurement techniques by combining similarities from multiple sources.

## B. Extrapolate Phases

Once the dataset is labeled, the next step is to extrapolate the labels from the specific compositions at which they were measured as shown in Figure 1d. To accomplish this, we fit a variational Gaussian process (GP) classifier to the labeled data. The details of our GP implementation and optimization process can be found in Section S1.2 of the Supplementary. From the optimized GP, we evaluate 2 functionals for each of the $N$ phases identified in the previous step: the mean, $\mu_i(x^*)$, which represents the probability of phase $i$ existing at any (measured or unmeasured) composition $x^*$ and the posterior uncertainty, $\sigma_i(x^*)$, which is the variance in $\mu_i$ at $x^*$. Using these functions, we can produce composition maps that, at every composition, identify the most likely phases and our overall confidence in that prediction.

## C. Choose Acquisition

The final analysis step of the agent is to choose the next sample composition for measurement that best accomplishes the campaign's goals. Examples of campaign goals include mapping all phase boundaries of a system, mapping those of a specific phase, or optimizing a physical property calculated from the SAS data or other measurement. In the active learning community, acquisition functions are typically described as combining the characters of exploitation (trusting and leveraging the current surrogate model) and exploration (searching outside of already sampled regions). For a typical task of identifying all phase boundaries, we choose to measure at the points of highest variance which is typically characterized as "pure exploration". The approach is grounded in the fact that the GP's uncertainty is maximized when the probability of multiple phases existing at a point is equal *i.e.* a phase boundary. We make three changes to the classic pure exploration approach: (1) We regularly measure poorly sampled regions in composition space, (2) we ensure that no measurements are chosen too close to one another, and (3) we randomly sample from the top 5 % to 10 % of variance values rather than choosing the singular true maximum value.

We choose to employ this "super exploration" since, in both *in silico and* experimental studies (*vide infra*), we found a series of edge conditions where the combination of the experimental data, classification parameters, and GP kernel tuning would produce what can be described either as a very stable GP solution with maximum uncertainty at a single point, or an insatiable drive to re-measure the same region of phase space. To mitigate this in the context of extended runs without human guidance, we incorporated a "periodic random step"; every $n$ iterations of the active learning loop (where $n$ is between 3 and 10), the input uncertainty function is replaced with a random field. As a result, due to the restriction on the closeness of measurements referenced above, the system samples a random region of phase space with low measurement density. We find this approach highly effective in avoiding local maxima of uncertainty and providing more reliable unattended sampling.

To further stabilize the system against such problems, we select our next measurement from the maximum uncertainty in a slightly unconventional way: we select all points from our fixed calculation grid with uncertainty within a certain percentage of the maximum, typically the top 5 % to 10 % of uncertainty, and then uniformly sample from this set. In doing this, we effectively buffer the measurement engine from numerical oddities of the GP solve, while still making quantitative use of the uncertainties generated.

This acquisition function approach has proven to be both stable and produce trusted results. Furthermore, it is highly adaptable and can be tuned for specific tasks (e.g., boundary identification of a specific phase or cost optimization) or to incorporate material non-idealities or instrumental effects (e.g., hysteresis or motor movement time). For the full details of our implementation, see Section S1.3 of the supplementary materials.

## III. *In silico* Agent Tuning

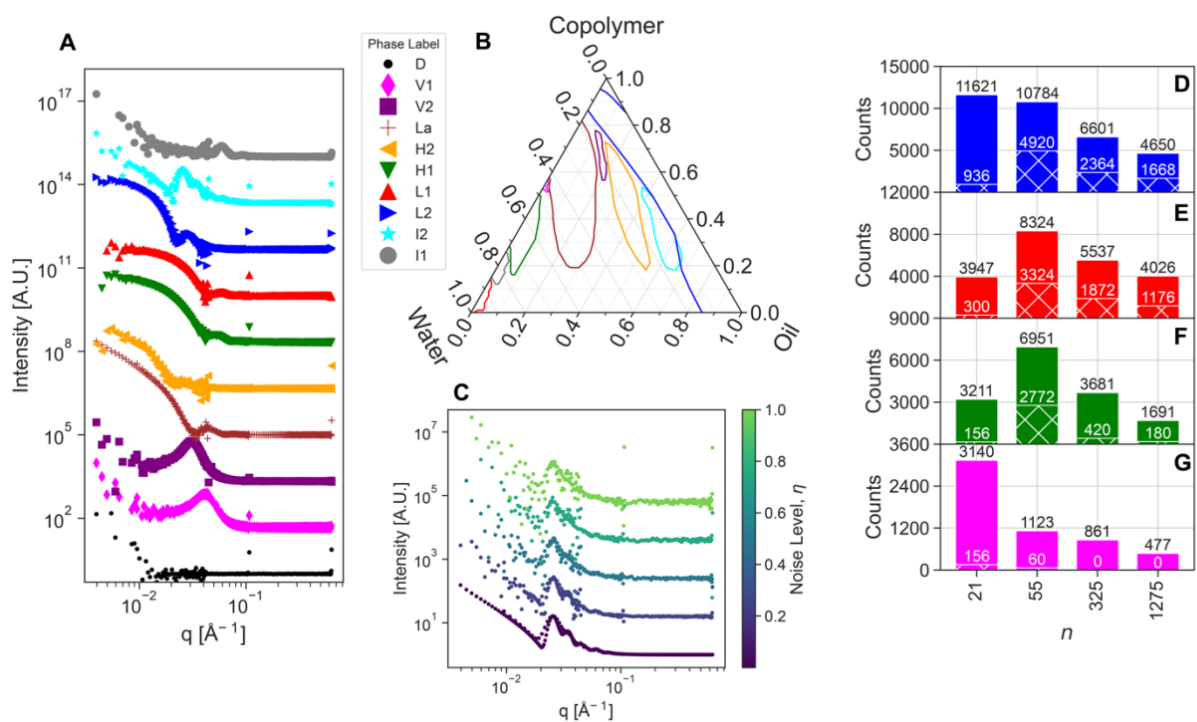

*Figure 2: Summary of in silico tuning results. (A) SAS data for each phase shown in (B) are generated using a scheme that incorporates instrumental smearing and tunable counting noise (C). (D-G) The number of labeling pipelines, out of >250,000 considered, which produce FMS ≥ 0.85 (D), 0.9 (E), 0.95 (F), and 1.0 (G) for varying number of measurements, $n$. The ground truth phase diagrams for each $n$ is shown in Figure S3. The hatched bars represent the number of pipelines that meet that subplot's FMS criteria at all noise levels rather than being considered individually.*

A central challenge in the application of autonomous approaches to experimental data is the availability of high-quality reference data with realistic artifacts, noise, and other features for training and benchmarking of classifiers, tuning of GP kernel parameters, and acquisition function selection. Before deploying the agent on the AFL platform, we tested, tuned, and benchmarked its performance against *in silico* experiments.

For these experiments, we sought to design a set of tests that would provide a significant challenge the agent and allow us to benchmark it in a challenging scenario. We identified the phase diagram in Reference [30] as having a large number of phases with complex boundaries, while also being of a material class directly relevant to the AFL program and our stakeholders. We chose analogous SAS models (Figure 2A) for each of the phases (Figure 2B) in Reference [30] and, to ensure that the tuned agent would be performant on real data, we introduced experimental artifacts into the synthetic data using real measurements as templates. This includes resolution smearing, statistical noise, and artifacts from stitching data from multiple instrument configurations together consistent with data from the 10m SANS instrument at the NIST Center for Neutron Research. See Supplementary Section S2 for the full details on this process. Furthermore, we introduce $q$-dependent counting noise into the data (Figure 2C) that is based on the reference measurements but can be amplified or diminished by a noise level scale factor, $\eta$. Given the phase boundaries and reference measurements, this virtual instrument will identify the ground truth phase based on requested composition and produce a scattering pattern that can be analyzed by the agent. In this way, our agent can query the virtual instrument, add the resulting scattering pattern to its measurement corpus, and then complete a virtual closed loop by predicting and requesting a measurement at the optimal next composition.

We break the agent tuning into two parts. Given the significant challenges associated with the labeling step of the agent, as discussed in Section II.A, we first focus on identifying a robust labeling pipeline. Our goal is to ensure that the labeling pipeline is robust at both small and large numbers of measurements, so we tested against symmetric grids with varying number of measurements, $n$, as shown in Figure S3. We then constructed over 250,000 labeling pipelines by varying both the clustering method and the form and coefficients of the similarity matrix calculation (See Supplementary Section S1.1.2). We quantify the performance of each pipeline using the Fowlkes-Mallows score defined as

$$FMS = \frac{TP}{\sqrt{(TP+FP)(TP+FN)}} \tag{1}$$

where TP refers to a true positive labeling, FP is false positive, and FN is false negative. $FMS$ varies between 0 and 1 with 1 being a perfect match to the ground truth and 0 being a perfectly incorrect labeling. Figure 2D-G shows the number of labeling pipelines that scored FMS ≥ 0.85 (D), 0.9 (E), 0.95 (F), and 1.0 (G) as a function of $n$. As expected, as we increasingly constrain the performance of the agent going from D-F, we see the number of pipelines that satisfy the constraint decreases. This is particularly true when we consider the agents performance across different noise levels $\eta$ (hatched bars) rather than individually (solid bars). For the $\eta$ constrained case, we see that no agents have a perfect score, FMS = 1.0, at all $n$ and all $\eta$. We find that the pipeline that has the highest FMS performance across all $n$ and $\eta$ has a minimum FMS of $\approx 0.946$ and consists of a Gaussian mixture model and sigmoid similarity. While this pipeline has the most robust performance, one might want to choose an agent that has a higher FMS for a given noise level and number of measurements. In Table S1 we show a list of high performing pipelines and their parameters.

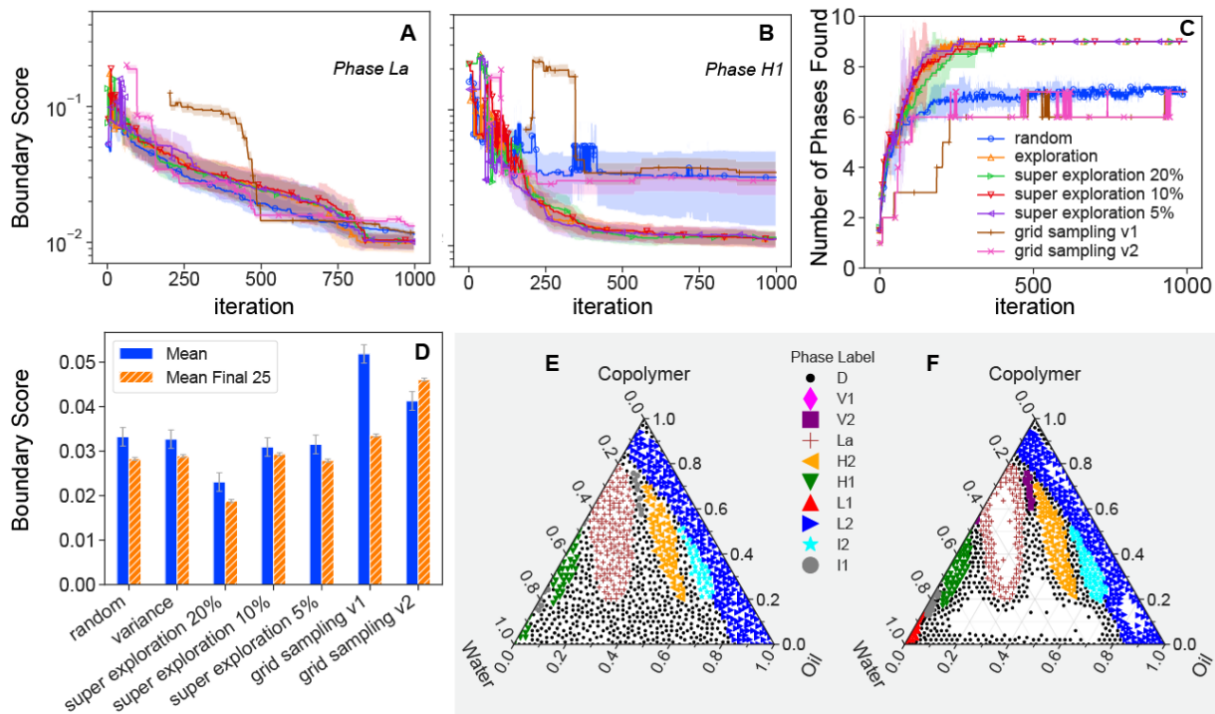

*Figure 3: Performance of the AFL agent with full in silico active learning runs with different acquisition functions as shown in the legend in Part C. Parts A and B show the boundary score as a function of iteration (number of measurements) for the (A) La, (B) H1, phases while Part C shows the number of correctly identified phases (out of 10). Details of the boundary score calculation can be found in Supplementary Section S3.4. Part D shows the average boundary score for each agent calculated over all or the last 25 iterations. For A-C, the lines and shaded regions represent the mean and standard deviation of seven independent active learning runs from different initial conditions. The error bars in Part D are the standard deviation across all trials, phases, and steps. Parts E and F show the final state of the phase map for random and exploration-based acquisition functions respectively.*

With the labeling step tuned for SANS measurements, we are ready to run full active learning campaign simulations. In Figure 3, we show the *in silico* performance of agents with different acquisition functions. Along with random-sampling and exploration (variance) based search, we show three versions of "super exploration" which use random instead of variance sampling for 5 %, 10 %, or 20 % of iterations as labeled. We also show two versions of multi-resolution grid sampling designed to mimic an experimentalist naïvely stepping through a preprepared set of samples. For grid sampling v1, we measure from 0 to 100 % water at 0 % copolymer, increase by a copolymer content by 5 %, and then repeat until we reach 100 % copolymer content. For grid sampling v2, we still measure from 0 to 100 % water at constant copolymer concentration, but, instead of a constant step size in copolymer concentration, we repeatedly bisect the copolymer composition space. More details of these agents can be found in Supplementary Section S3.5. We quantify the performance of our agents both in terms of the "boundary score" (Figure 3A, 3B) and the number of phases identified by the agent (Figure 3C). The boundary score is the average distance of an agent identified boundary from its ground truth boundaries and is described in detail in Supplementary Section S3.4. We use the boundary score rather than calculating the FMS at each point in a composition grid, as the latter approach would bias the performance evaluation towards the largest phases by area. Furthermore, our boundary score metric is better aligned with our agent goal which is to accurately identify the location of phase *boundaries.* Figure 3A and 3B display the boundary as a function of iteration (the number of measurements) for two representative phases within the challenge problem. Note that all phases are mapped simultaneously despite each plot showing the performance of mapping a single phase. The boundary score plots for all phases are shown in Figure S6. Figure 3D also shows, for each agent, the average boundary score calculated over all iterations and just the final 25 iterations. Respectively, these derived values characterize the speed at which each agent correctly finds the boundary and the final accuracy of the agent identified boundaries.

In all cases, the exploration-based agents show superior performance to the random sampling or grid searches. Figure 3C shows that these agents locate more phases on average and, Figure 3D shows that the super exploration agent with 20 % random sampling is superior to all other agents in both speed and accuracy. While random sampling initially reduces the boundary score more quickly than the exploration-based agents for the largest phases (Figure 3A), the exploration-based agents show the greatest improvement at the end of the campaign and significantly better performance for the smaller phases. For the random and grid-based searches, the lack of data clustered around small phases leads to a breakdown in the labeling performance, leading to incorrect or non-identification of phases. Finally, it's clear that the exploration-based agents spend more time measuring near phase boundaries than the random search from the ternary diagrams in Figure 3E and 3F respectively.

These *in silico* tests and tuning runs are crucial towards making effective use of limited neutron or X-ray beamtime at a user facility. In addition, they also validate the agent's design and show that it reduces the number of measurements needed to map a phase space. A key feature of this sampling approach is that we identify the location of phase boundaries with far greater accuracy and greatly reduce the number of points needed to map a phase space when compared to a naïve grid search.

## IV. *In operando* Demonstration

After demonstrating the AFL agent *in silico*, the next step is to validate the same agent in a live SAS experiment. This validation will not only verify the efficacy of the agent in mapping phase boundaries, but also the *in silico* approach used to tune the design and parameters of the agent. The latter is particularly important given the limited availability of neutron and synchrotron X-ray scattering beamtime. While we expect that the parameters we identified to be general, we also recognize that some experiments may require specific tuning of the agent in order to achieve optimal performance. Based on this reasoning, the following results are not intended to be a measure of top performance of the agent but rather a validation of using *in silico* tuning to achieve a reasonably performant agent with minimal excess beamtime used for tuning.

The agent is implemented in our open-source software package, AFL-agent.[31] The agent is deployed to the AFL platform [4] as a microservice (HTTP service local to the instrument) using the AFL-Automation APIServer architecture. This "*AgentServer*" interfaces with a custom-designed sample server (*SampleServer*) to facilitate closed-loop experimental measurements. The *AgentServer* handles the labeling, extrapolation, and acquisition function steps while the *SampleServer* is responsible for orchestrating the preparation, measurement, and cleanup of a sample. Separating the guidance agent from the *SampleServer* allows rapid debugging and tuning of agent hyperparameters such as similarity functions, GP kernel parameters, and acquisition functions during a run without interrupting measurement, which can be particularly important for long running measurements in, e.g., neutron scattering.

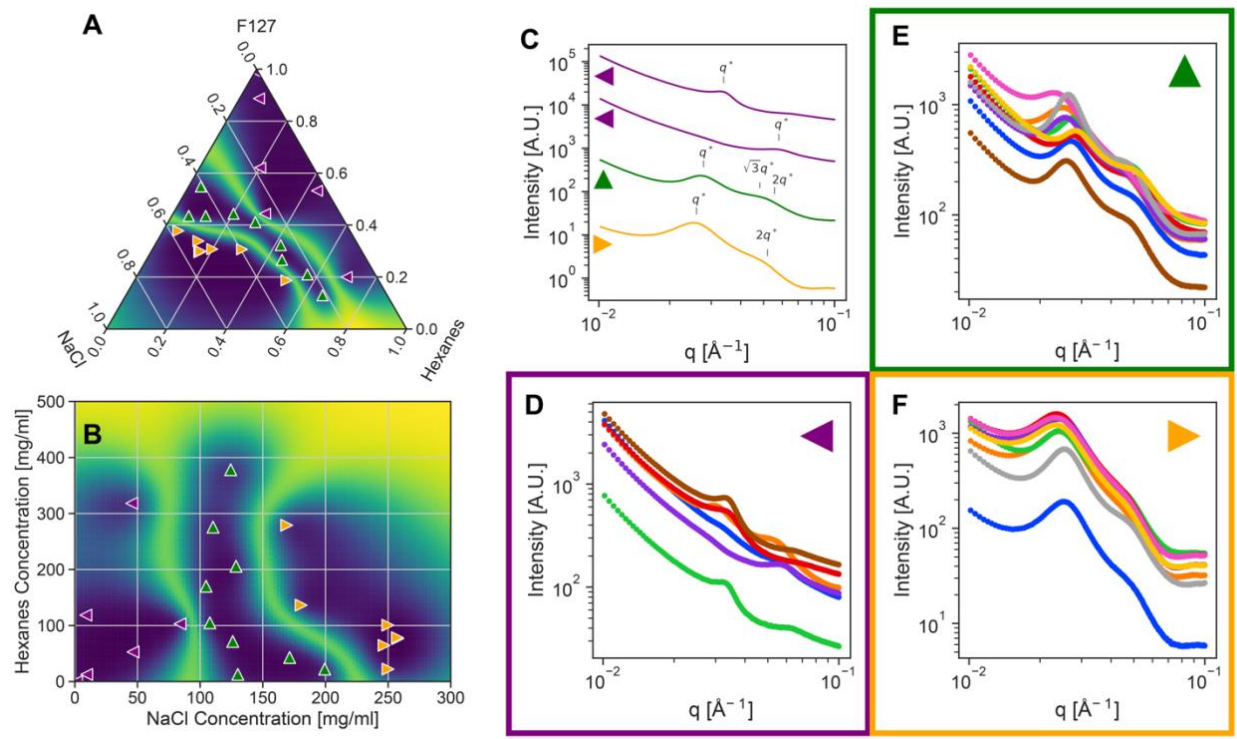

*Figure 4: Experimental active learning results on mixtures of Pluronic F127, NaCl, hexanes, and water. The final ternary phase diagram measured by the agent and (B) the binary projections of this phase diagram with the symbols denoting the phase identified by the agent and the background colormap the uncertainty in this phase assignment. Parts D-F show the SAS data associated with each agent identified phase. Part C shows four representative datasets with the characteristic diffraction peaks labeled indicating, from top to bottom, spherical micelles, spherical micelles, hexagonally packed cylinders, and lamellar structures.*

As a starting point we use X-ray scattering to map the phase behavior of a non-ionic copolymer surfactant formulation analogous to the one studied *in silico*: Pluronic F127[1], salt (NaCl), hexanes, and water. In this experimental campaign, we varied the mass fractions of F127, NaCl, and hexanes and fixed the total volume of each sample, thereby constraining the water component of the mixture. The results are shown in the ternary diagram in Figure 4A and the projection of the ternary into a binary concentration space appear in Figure 4B. The uncertainty colormap in these figures serves as a proxy for the location of the phase boundaries, with the brighter colors that indicate higher uncertainty in phase label also indicating the likely location of the structural phase transition. The agent identifies three clusters in the SAS data and we've labeled the identifying structure factor peaks on representative data from each of these clusters in Figure 4C. At low salt concentrations there are two populations of micelles of different radius (Figure 4D) which, as the salt concentration is increased, transition to what is consistent with hexagonal ordering (Figure 4E) and then lamellar ordering (Figure 4F). While the dependence of salt on the phase behavior of non-ionic surfactants has been reported previously,[32] our agent has re-discovered this trend independently, without prior or programmed knowledge of this phenomena.

---

[1] Certain commercial equipment, instruments, or materials (or suppliers, or software, …) are identified in this paper to foster understanding. Such identification does not imply recommendation or endorsement by the National Institute of Standards and Technology, nor does it imply that the materials or equipment identified are necessarily the best available for the purpose.

It is important to highlight that the AFL agent performed well in this experiment without ever having been exposed to scattering data of this form. Despite our extensive attempts to benchmark the agent with "realistic" virtual instrument data, it's clear that none of the scattering in Figure 4 look quite like the data in Figure 2A. The experimental data has a strong background signal and displays Bragg scattering consistent with crystalline materials rather than the primarily "form factor" scattering that was used for the *in silico* testing. Regardless, our similarity approach to identifying data clusters allowed the agent to find regions of similar scattering within the explored composition space. While a human practitioner might have separated the two micelle structures into different phases because their sizes are different, it is entirely reasonably to construct a taxonomy that gathers the "single particle" scattering into a single grouping.

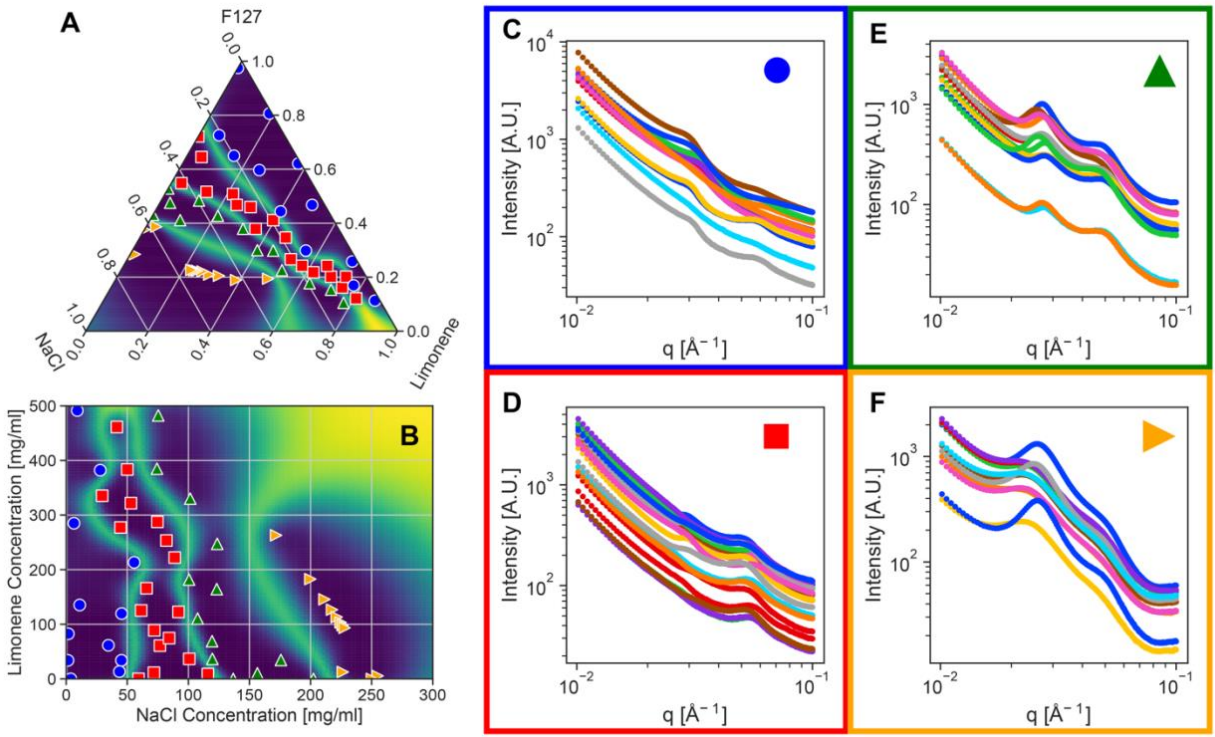

*Figure 5: Experimental active learning results on mixtures of Pluronic F127, NaCl, hexanes, and water. (A) The final ternary phase diagram measured by the agent and (B) the binary projections of this phase diagram with the symbols denoting the phase identified by the agent and the background colormap the uncertainty in this phase assignment. Parts C-F show the SAS data associated with each agent identified phase.*

As an example of a product reformulation challenge, we took the same formulation discussed above and replaced the petroleum derived hexanes with limonene, a naturally derived analogous oil. The promise of autonomous exploration is that such challenges can be accelerated and the differences in phase behavior can be rapidly identified. We performed the same campaign with the same agent, simply replacing the component. The results are shown in Figure 5. In contrast to the hexanes system, the agent separates the micelle scattering into two groupings in the limonene system. This difference in classification likely originates from the increased number of datapoints we were able to gather for the limonene case. The analogous phase to the hexagonal phase observed in the hexanes case begins at a similar salt concentration but is a stronger function of the oil component concentration, with salt and limonene having a nearly equal role in driving crystalline structure formation. Similar to the hexanes system, the hexagonal phase appears at ~100 g/ml NaCl and transitions to lamellar ordering at higher salt concentrations. The width of the identified phase boundaries and their multidimensional shape are critical variables for designing liquid formulations with the desired stability or engineered transitions.

For these solutions, unraveling the phase boundary's co-dependence on limonene and salt using conventional scattering experiments with linear grid spacing would be challenging. In the worst-case scenario, we estimate that the grid sampling approach would require as many as 25 x more measurements to achieve the same resolution in boundary location. Furthermore, our *in silico* tests show that the grid approach can miss phases (Figure 3D) and generally poorly identifies the location of boundaries. GP-driven sampling provides a significant improvement over grid sampling as the boundary is only constrained by the Gaussian process kernel and more measurements are conducted near the phase boundaries.

The datasets shown in Figure 4 and Figure 5 exemplify the challenges of studying soft-material phase spaces. While there are similarities to the two systems, the reality of the formulations industry is that precise knowledge of system behavior at specific concentrations, temperatures, and processing conditions is crucial for manufacturability. Manufacturers often want to minimize or maximize certain components to minimize product cost and this can only be achieved with precise knowledge of a phase boundary. Active learning tools, like the one described in this paper, allow industry to rapidly and efficiently map phase spaces.

## V.   Conclusion

We have described the design, tuning, and validation of an active learning agent that maps the phase spaces of soft-material formulations using SAS. The agent is designed to be general and does not require *a priori* knowledge of the phase behavior of a system under study. The agent is modular so that, in future studies, it can be optimized for other instruments, to include multiple measurement modalities at once, or to include instrument non-idealities not considered in this work. We tuned the agent through extensive *in silico* experimentation using synthetic data that includes real instrument artifacts including *q*-dependent counting noise, resolution smearing, and stitching artifacts. We showed that this tuning resulted in an agent that is robust to measurement noise and is effective at both small and large numbers of measurements. Finally, we applied the tuned agent to a model formulation problem where we compare the phase maps of two formulations where the difference is whether a petrol- vs naturally derived oil is included in the formulation. The agent primarily identifies a crystallization transition in these formulations, and we show that, through post-hoc analysis, the agent could be tuned to identify several unique non-crystalline morphologies if that was the focus of a formulation study.

Active learning agents like the one described here promise to not only revolutionize the way we do measurements but, more broadly, the way we design product formulations. In our current Edisonian world, manufacturers spend significant capital to produce singular, high performance product formulations that minimize production cost, meet regulatory frameworks, and meet the needs of consumers. A highly tuned autonomous platform, such as the AFL, promises to greatly reduce the reformulation time of products leading to a greater number of better, greener choices for consumers. Taking this concept a step further, a fleet autonomous platforms deployed to various local manufacturers could map and optimize hyper-local products that use dynamically or seasonally varying feeds streams. While there is significant work to be done towards achieving this goal, our AFL platform advances towards a future with greener, cleaner formulations.

## VI. Acknowledgements

Partial support for this work was provided by the members of the *n*Soft industrial consortium (nist.gov/nsoft). This work is based on research sponsored by AFRL under agreement number FA8650-19-2-5220. The U.S. Government is authorized to reproduce and distribute reprints for Governmental purposes notwithstanding any copyright notation thereon

*Supplementary Information for*

# Autonomous Small-Angle Scattering for Accelerated Soft Material Formulation Optimization

Tyler B. Martin[*], Duncan R. Sutherland Austin McDannald, A. Gilad Kusne, Peter A. Beaucage[*]

1. National Institute of Standards and Technology, Gaithersburg MD, 20899

* corresponding authors: tyler.martin@nist.gov and peter.beaucage@nist.gov

# Table of Contents

## VIII. Details of the AFL Agent

The processing steps of the AFL agent are schematically shown in *Figure S6* and are discussed in detail in the following paragraphs. As our agent relies heavily on open-source implementations of various machine-learning methods, where appropriate, we will broadly describe our usage of a method and provide a reference for the full implementation details of the specific version of the software we used. Our agent codebase `AFL-agent` is open-source and can be downloaded from our usnistgov GitHub repository.[1]

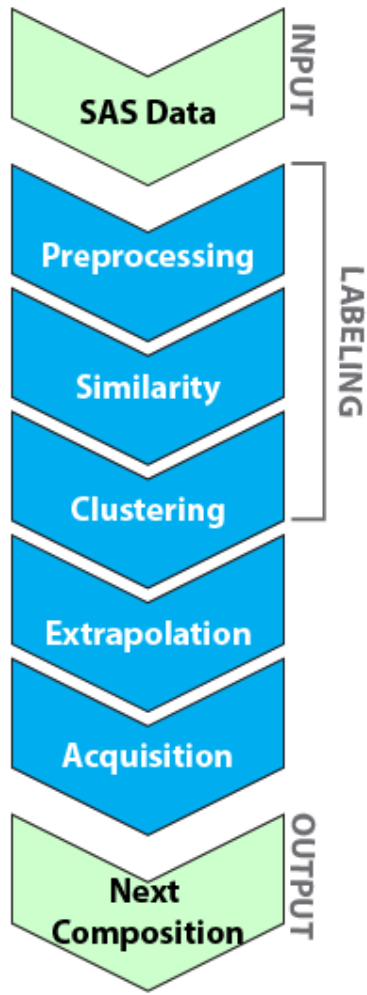

*Figure S6: Schematic description of the data processing steps of the AFL agent.*

## A. Labeling

In this work, the labeling step is composed of a three-step pipeline. The advantage of our labeling approach is that it is very general and can be applied to many material systems and measurements. The downside is that this approach can struggle to distinguish subtle changes in the measurement data or handle continuously changing (2$^{nd}$ order) phase boundaries. Other labeling strategies such as classification provide an attractive alternative but require large amounts of high-quality labeled data. Applying classification approaches to our pipeline is the subject of future work within our program.

### 1. Data Preprocessing

The first step in the agent pipeline is to preprocess, normalize, and correct the provided small-angle scattering (SAS) data. This step is to ensure that the agent can be performant across a variety of materials and instruments. The data is first trimmed to a $q$-range (domain) that is (a) appropriate for the SAS instrument and (b) contains the feature of interest in the scattering pattern. For the latter, the $q$-range can be trimmed or segmented to remove features not associated with the physical phenomena of interest in a given study. Next the dataset has all NaN ("not a number") values removed and is back-filled and forward-filled with constant values so that all data is of the same length. This filling of the data is to account for the use of data from different instruments *i.e.* if reference data from one instrument is used to seed a campaign on a different instrument with a slightly different q-range. This is important as some instruments will drop $q$-values with low signal-to-noise commonly seen at the extrema of a scattering range. Finally, the data is linearly interpolated onto a common, geometrically spaced q-grid. A Savitsky-Golay (SG) filter is applied to lightly smooth each pattern and then calculate the 1$^{st}$ and 2$^{nd}$ derivative.[2] For the SG processing, we typically use a window length of 31 points and a polynomial order (polyorder) of 2.

*2.    Similarity*

Next is the similarity calculation. Here, we choose from several similarity metrics and coefficients identified in our *in silico* analysis discussed in Section III of the main text. The most common metrics we use are the Laplacian kernel

$$W(I_i(q), I_j(q)) = exp\left(-\gamma \|I_i(q) - I_j(q)\|_1\right) \quad (1)$$

and the polynomial kernel:

$$W(I_i(q), I_j(q)) = \left(\gamma I_i^T(q) I_j(q) + c_0\right)^d \quad (2)$$

Where $I_i(q)$ and $I_j(q)$ are the scattering intensities of measurements $i$ and $j$ as a function of $q$, $\gamma$ & $c_0$ are coefficients used to tune the kernel and $d$ is the power law exponent. We choose a similarity function and apply it to all pairs of data and construct a matrix of similarity values. For all *in silico campaigns* in the main text we used a Laplacian kernel with $\gamma = 0.0025$ applied to both the SG filtered data (0[th] derivative) and the 1[st] and 2[nd] derivatives. We then sum the two similarity matrices and normalize them using the relation:

$$W(I_i(q), I_j(q)) = \frac{W(I_i(q), I_j(q))}{\sqrt{W(I_i(q), I_i(q)) W(I_j(q), I_j(q))}} \quad (3)$$

The similarity calculation for the experimental validation was identical except the 2[nd] derivative was omitted.

*3.    Clustering*

With the normalized similarity matrix calculated, we next apply a clustering routine to gather the data into groups. As discussed in Section III of the main text, we have focused on two clustering methods: *spectral clustering* and *Gaussian mixture models*, as implemented in version 1.3.0 of scikit-learn.[3, 4] We use the default parameters with both of these methods except for setting `affinity="precomputed"` for spectral clustering.

For both clustering methods above, the number of clusters (or phases, $N$, in our case) must be specified before running the routine. To determine the number of phases, we use a modified silhouette score method as implemented in scikit-learn version 1.3.0.[5] In this method, the Silhouette Coefficient for measurement $i$ is calculated as

$$s_i = \frac{b-a}{\max(a,b)} \tag{4}$$

where $a$ is the average similarity between all of measurements that measurement $i$ belongs to and $b$ is the average similarity between measurement $i$ and the next closest cluster of measurements. To choose the optimal number of clusters, we repeat the clustering and $s_i$ calculations for $N = 2 \ldots 10$ clusters and calculate the mean Silhouette Coefficient, $\bar{S}$, for each. Values of $\bar{S}$ close to 1 indicate a high confidence in the clustering while lower values (with a minimum of 0), indicate lack of confidence in the clustering for that number of clusters. Rather than taking the clustering corresponding to the maximum $\bar{S}_N$, we choose the clustering with the largest $N$ that has $\bar{S} > 0.85$. If no clustering satisfies this constraint, the constraint is reduced by 0.05 until a clustering satisfying the constraint is found. If no cluster has $\bar{S} > 0.4$, then we assume that $N = 1$. In our testing, this heuristic approach stabilizes the prediction of the number of phases and produces prediction more consistent with human intuition.

## B. Extrapolation

Here, we employ a variational Gaussian process (VGP) classifier as implemented in scikit-learn v1.3.0 and GPFlow version 2.9.0 for the virtual testing and experiments respectively.[6, 7] The VGP allows us to use a non-Gaussian likelihood which is necessary for the implementation of a muti-class classifier.

For the GPFlow implementation of the VGP, we use the Matern32 kernel, RobustMax link function and a MultiClass likelihood.[8-10] The kernel was chosen via an abbreviated version of the *in-silico* tests described in the text, while the link and likelihood functions are recommended choices from the GPFlow documentation. The VGP is fit to the results of the clustering step with the input data being ternary compositions and the output being the phase label. Once the optimization is complete, the VGP can then predict the mean, $\mu_i(x^*)$, which represents the probability of phase $i$ existing at composition $x^*$ and the posterior uncertainty, $\sigma_i(x^*)$, which is the uncertainty in $\mu_i$ at $x^*$.

For the scikit-learn implementation of the VGP, we used a Matern kernel with `nu=1.5` and an initial length scale of `length_scale=1`. When using this implementation, we use entropy, $E$, as a stand-in for variance which we calculate from the mean functions for phase $i$, $\mu_i$, as

$$E = -\sum_i \mu_i \log \mu_i \tag{5}$$

## C. Acquisition

Finally, the results of the VGP calculation can be used to choose then next sample to prepare and measure. For the purposes of phase mapping, we use a variance based 'super exploration' acquisition function which modifies the traditional 'pure exploration' approach. The choice of exploration based acquisition works in this case as, due to the construction of the VGP, the uncertainty is guaranteed to be maximized when the probability of multiple phases existing at a composition are equal (*i.e.,* a phase boundary). For pure exploration, we would take the posterior uncertainties calculated from the VGP, sum them to create an overall uncertainty, and then choose the next composition at the point of maximum overall uncertainty. For 'super exploration', we modify the way the composition is chosen from the uncertainty in two ways. First, rather than choosing the maximum uncertainty, we randomly choose a composition from the highest 3-5 % of the uncertainty distribution. Secondly, we introduce a constraint that our selected point cannot be within 1.5 % of an already measured point. These two modifications help alleviate issues with the agent getting 'stuck' and oversampling portions of the phase diagram. Finally, every $n_{density\_sample}$ active learning iterations we switch to an acquisition function that samples based on the point density of measurements rather than the uncertainty. Specifically, we fit a Kernel Density estimation model [11] to the ternary composition values and use this to calculate the log-likelihood of having measured at given composition. From this, we sample the composition space and randomly choose a position that has a low likelihood, which corresponds to an undersampled portion of the phase diagram. This final modification ensures that we don't place too much trust in the VGP model and that we have measurements that span the available composition range.

## IX. Details of Synthetic Data Generator

Our synthetic SAS data generator is built up in two steps. First, a set of compositions and phase labels is gathered from an experiment or by manually tracing the phase boundaries of a figure from the literature. For each unique phase, the alpha shape of the set of points is calculated using version 1.3.1 of the alphashape package.[12] Using this tool, the boundary (*concave hull*) of each phase can be identified for visualization. Most importantly, given an arbitrary point in our composition space, we can iterate over the alpha shapes and find the phase identity of the point.

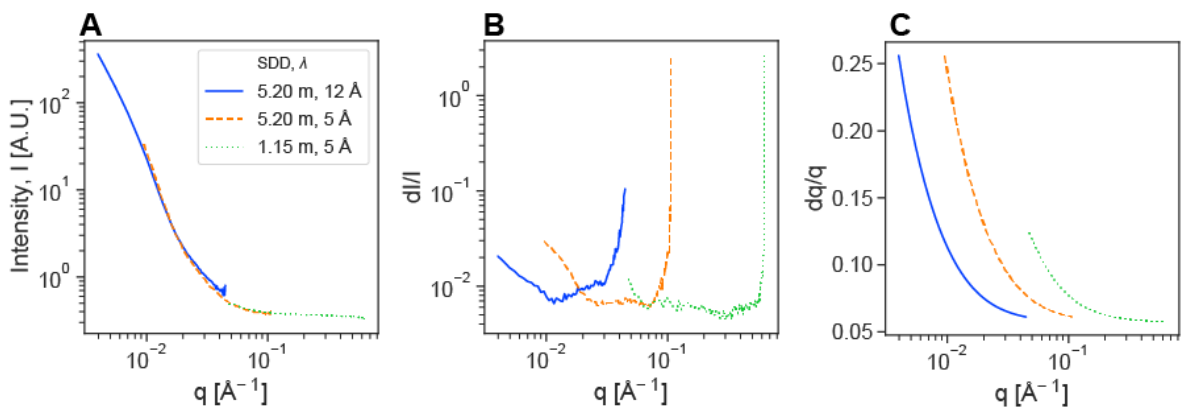

*Figure S7: Reference SAS measurements used for introducing resolution, smearing, and counting noise into generated model data. From the measurements, we show the (A) scattering intensity, (B) intensity normalized Poisson noise, $\frac{dI}{I}$, from the 2D area detector propagated through an azimuthal integration, and (C), the q-resolution function normalized by the q-values, $\frac{dq}{q}$. The data were collected on the 10m SANS instrument at the NIST Center for Neutron Research in three configurations corresponding to two sample to detector distances $SDD = 5.2\,m$ and $1.15\,m$ and two wavelengths $\lambda = 12\,\text{Å}$ and $5\,\text{Å}$. The data correspond to a suspension of perfluoronated polymer and carbon nanoparticles. We note that the material identification is done only for completeness as the purpose of these data are just to quantify the instrumentation resolution functions and counting statistics.*

The second step is to build the SAS generator for each unique phase. Our approach is to take an analytical model and add in resolution smearing and $q$-varying counting noise using experimental reference measurements. Our reference data is shown in Figure S7. For the discussion in the main text, our analytical models are chosen from version 1.0.7 of the sasmodels package,[13] although any SAS model generator could be used. The data is resolution smeared by taking the resolution function from reference measurements and convolving it with the model data as described in the SasView documentation.[14] The variable counting uncertainty for each $q$ is calculated by drawing random values from a normal distribution with the mean taken as the measured scattering intensity $I_{expt}(q)$ and the uncertainty $\sigma_{synthetic}$ defined as

$$\sigma_{synthetic}(q) = \sigma_{expt}(q) \left( \frac{\eta}{\left(\frac{1}{N_q}\right) \Sigma_q \left(\frac{\sigma_{expt}^2(q)}{I_{expt}(q)}\right)} \right) \tag{5}$$

In this expression, $\sigma_{expt}(q)$ is the q-dependent noise calculated initially as Poisson noise on the 2D area SAXS or SANS area detector and propagated through the azimuthal integration. The term $\eta$ is a tuning parameter used to control the level of counting noise in the synthetic data and $N_q$ is the number of q values in the 1-D dataset.

The above approach is an attempt at introducing experimental effects into theoretical models in an efficient, tunable way that avoids expensive Monte Carlo simulations. The inclusion of experimental resolution functions ensures that we don't optimize our agent for unobtainable features of scattering models (e.g., perfect Bessel function fringes). The tunable noise allows us to simulate undercounted measurements which, for the purposes of autonomous learning, might be preferred to maximum the number of compositions that can be sampled. Furthermore, for SAS instruments where data is stitched together from multiple instrument configurations, as is the case with most SANS instruments that are not time-of-flight based, our approach can be iteratively applied. In this mode, the resolution function and counting noise from each configuration is used to generate a synthetic scattering pattern and then these patterns are stitched together. Importantly, theses stitched curves have the same stitching artifacts that would be present in the real data making our tuning even more relevant to the experimental case.

## X. *in silico* Agent Testing

### A. Testing Datasets for Labeling Pipeline

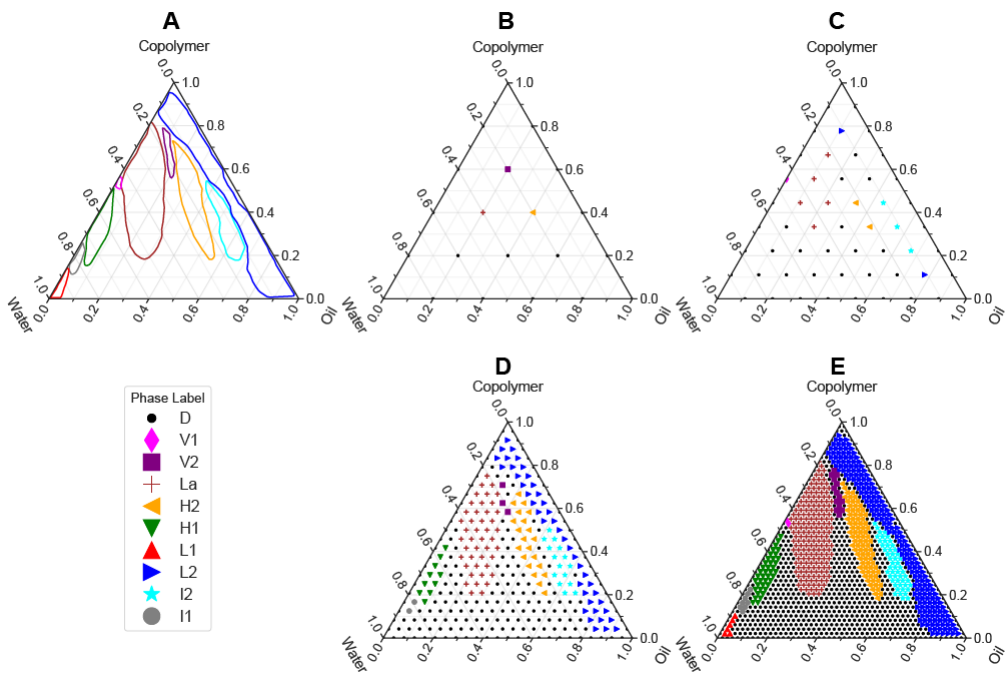

*Figure S8: (A) Full phase boundaries and (B-E) discrete datasets for in-silico labeling pipeline testing. Parts (B-E) correspond to $n = 21, 55, 325, and\ 1275$ measurements respectively. The colors and symbols match the legend between Figure 2A and Figure 2B of the main text.*

**B. Labeling Pipeline Performance with Spectral Clustering**

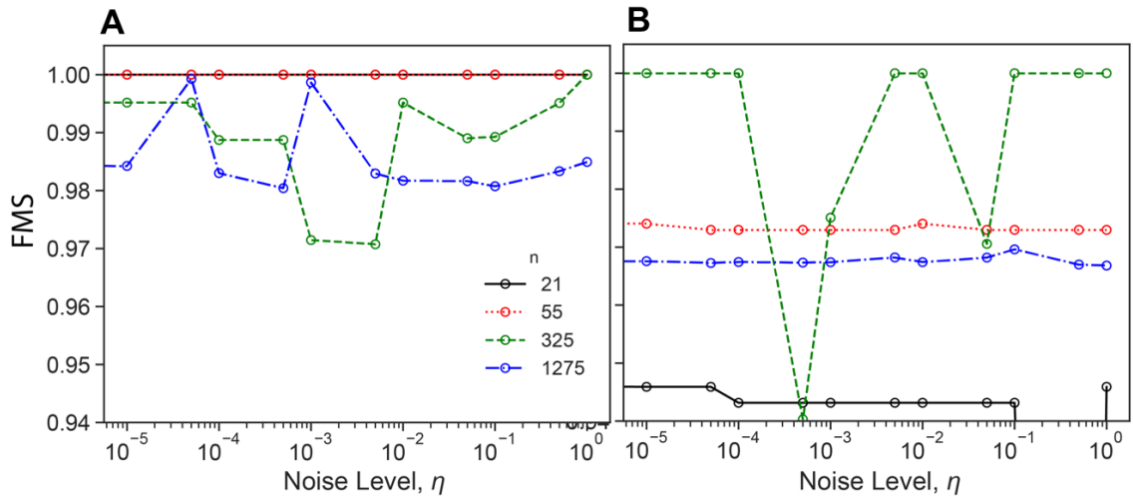

*Figure S9: Noise sensitivity of the best labeling pipelines for pipelines using (A) gaussian mixture models or (B) spectral clustering for the labeling step as a function of the noise level $\eta$ and number of measurements $n$.*

## C. Best Labeling Pipelines

*Table S1: Table of the best labelling pipelines identified by in-silico sampling.*

| $n$ | Clustering Method | Affinity Metric | Distance Matrix? | gamma | degree | c0 | co_gamma | FMS |
|---|---|---|---|---|---|---|---|---|
| 21 | gaussian_mixture_model | laplacian | TRUE | 0.00001 | N/A | N/A | 0.0001 | 1 |
| 21 | gaussian_mixture_model | laplacian | TRUE | 0.0001 | N/A | N/A | 0.001 | 1 |
| 21 | gaussian_mixture_model | laplacian | FALSE | 0.001 | N/A | N/A | N/A | 1 |
| 21 | spectral_clustering | laplacian | FALSE | 0.00001 | N/A | N/A | N/A | 0.9170 |
| 21 | spectral_clustering | laplacian | TRUE | 0.00001 | N/A | N/A | 0.00001 | 0.9219 |
| 21 | spectral_clustering | rbf | TRUE | 0.001 | N/A | N/A | 0.001 | 0.9233 |
| 55 | gaussian_mixture_model | laplacian | FALSE | 0.00001 | N/A | N/A | N/A | 1 |
| 55 | gaussian_mixture_model | laplacian | FALSE | 0.001 | N/A | N/A | N/A | 1 |
| 55 | gaussian_mixture_model | laplacian | TRUE | 0.001 | N/A | N/A | 0.00001 | 1 |
| 55 | spectral_clustering | poly | FALSE | 0.0001 | 2 | 100 | N/A | 0.9732 |
| 55 | spectral_clustering | poly | TRUE | 0.001 | 2 | 1000 | 0.00001 | 0.9732 |
| 55 | spectral_clustering | poly | TRUE | 0.001 | 2 | 1000 | 0.0001 | 0.9732 |
| 325 | gaussian_mixture_model | poly | FALSE | 0.00001 | 1 | 0 | N/A | 0.9813 |
| 325 | gaussian_mixture_model | poly | TRUE | 0.00001 | 1 | 0 | 0.00001 | 0.9815 |
| 325 | gaussian_mixture_model | poly | FALSE | 0.0001 | 0.5 | 1000 | N/A | 0.9899 |
| 325 | spectral_clustering | laplacian | TRUE | 0.05 | N/A | N/A | 0.05 | 0.9905 |
| 325 | spectral_clustering | poly | FALSE | 0.00001 | 2 | 10 | N/A | 0.9761 |
| 325 | spectral_clustering | poly | FALSE | 0.00001 | 4 | 100 | N/A | 0.9754 |
| 1275 | gaussian_mixture_model | poly | FALSE | 0.00001 | 1 | 1 | N/A | 0.9869 |
| 1275 | gaussian_mixture_model | poly | FALSE | 0.00001 | 1 | 1000 | N/A | 0.9774 |
| 1275 | gaussian_mixture_model | poly | TRUE | 0.00001 | 1 | 0 | 0.00001 | 0.9765 |
| 1275 | spectral_clustering | poly | FALSE | 0.00001 | 3 | 100 | N/A | 0.9676 |
| 1275 | spectral_clustering | poly | FALSE | 0.00001 | 4 | 100 | N/A | 0.9677 |
| 1275 | spectral_clustering | poly | FALSE | 0.0001 | 4 | 1000 | N/A | 0.9676 |

*Table S1* shows the results of our brute force sampling of labeling pipelines. From left to right the columns are described as follows. "$n$" is the number of measurements used in the test and corresponds to *Figure S8*. "Clustering Method" and "Affinity Metric" are the scikit-learn methods used in the clustering and similarity calculation steps. The "Distance Matrix?" column indicates whether the similarity matrix was multiplied (elementwise) with a Euclidean distance matrix between the compositions of the measurements. The goal was to introduce a locality to the clustering and biases against clusters that span the composition space. "gamma", "degree", & "c0" are the parameters used in the similarity calculation and "co_gamma" is the scaling factor used for the distance matrix (where used). Finally, "FMS" is the Fowlkes-Mallows score as described in the main text.

## D. Boundary Score Description

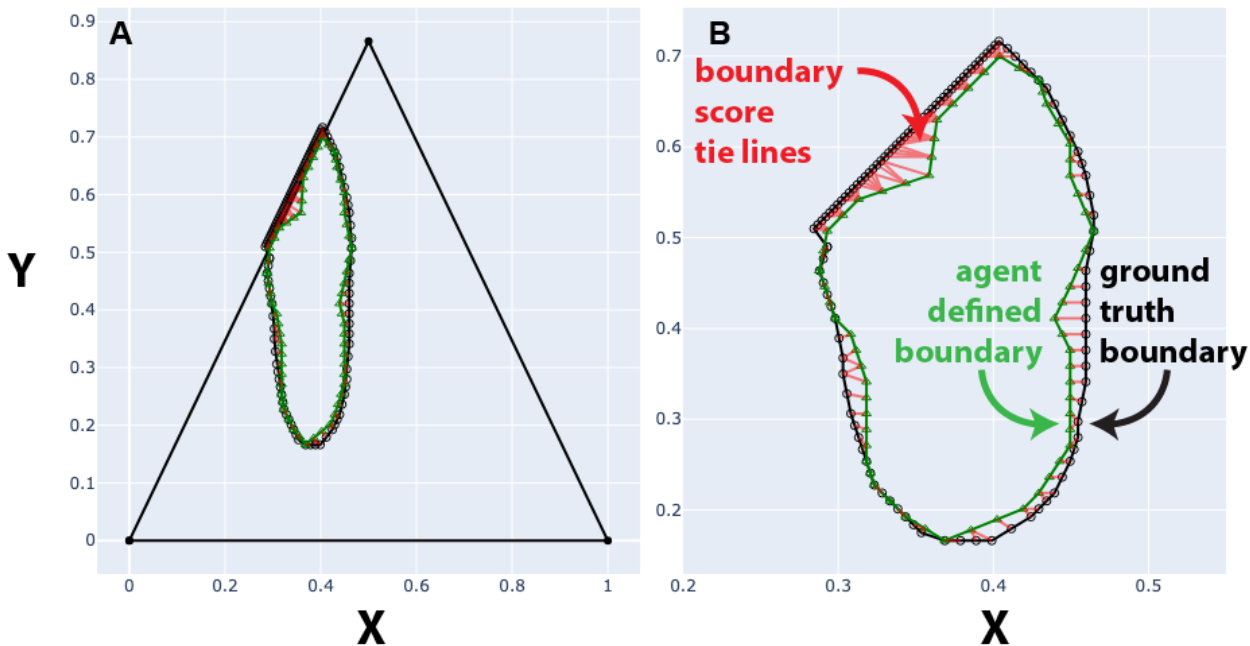

*Figure S10: Depiction of the boundary score calculation for the La phase (A) inside the full ternary and (B) with a magnified and labeled view. Both plots are in cartesian coordinates.*

A key challenge in benchmarking the performance of the AFL active learning agent is finding an appropriate metric quantify it. Traditional metrics for classification, such as the Fowlkes-Mallows score discussed in the main text, can be applied point by point on a composition grid but this approach is inherently area-biased. This means that the agent can completely misclassify small phases and still produce a high score if the largest phases are mostly correctly identified. Since the primary goal of this agent is to accurately identify the location of phase *boundaries*, rather than identifying phase *regions*, we needed a metric that focused on boundary location rather than area identification.

To address this, we have developed a metric which we call the "boundary score". This metric is calculated as the average distance between the ground truth phase boundary and the agent defined boundary for a given phase in a phase diagram. Briefly, each boundary is found by finding the concave hull for both the ground truth and the agent defined phase regions. Then, for each boundary point in both hulls, the closest boundary point in the opposite hull is found. The average distance between these points of closest approach between the hulls is the boundary score. In Figure S10, the boundary score can be visually understood as the average length of the red tie lines. From this, it should be clear that this metric is explicitly focused on the accuracy of phase boundary placement.

To outline the procedure for calculating the boundary score in more detail: The ground truth boundaries are found by hand-labeling a 5050 point grid of data from the phase diagram found in Reference [15]. Included in these 5050 points is a row of points outside of each side of the ternary boundary. These outside points are forbidden from being selected by the agent during the simulated campaign but are necessary for our "boundaries" to extend to the edges of the ternary during the next step. After the labeled grid is created, we find the boundary definitions using a concave hull calculation as implemented in the `shapely.concave_hull` method with `ratio=0.2`.[16] We then use the `shapely.segmentize` method with `max_segment_length=0.025`.[17] This creates a closed, dense set of points (black circles in Figure S10) which define the phase boundary.

The agent defined phase boundary (green triangles in Figure S5) is calculated similarly. The phase regions are identified by calculating the most likely phase at every composition, $x^*$, via the VGP derived mean function, $\mu_i(x^*)$ function, described in Section VIII. Once the phase regions are identified, we use the same concave hull approach described above to identify the boundary points of each phase.

We now calculate the boundary score between *each pair* of ground truth and agent defined boundaries. This means that *each* agent defined boundary will have $M$ scores corresponding to the $M$ ground truth phases. This is necessary because our clustering approach produces numerical labels that are unrelated to the physical ground truth labels. Therefore, we must score each agent-defined boundary against all ground-truth boundaries in order to conduct a quantitative matching process. For each boundary node in a given pair of ground truth and agent boundaries, we find the shortest distance to a node in the opposite boundary. After removing any repeated node pairs from the list, the average of these minimum distances is the boundary score for that pair of agent-defined and ground truth boundaries.

---

**Algorithm S1:** *Pseudo-code describing how the ground truth label of agent-identified phase boundaries are identified from boundary scores.*

---

1:    $n \leftarrow$ numerical label of agent identified phases
2:    $m \leftarrow$ phase label of ground truth phases
3:    $B \leftarrow$ list of all agent and ground truth boundary label pairs sorted by boundary score (lowest first)
4:    $P \leftarrow$ empty list to hold agent and ground truth phase boundary pairs
5:    **for each** pair of boundary labels (m, n) in B **do**
6:      **if** m is not in any pair in P **do**
        **if** $n$ is not in any pair in P **do**
7:          $P \leftarrow append(m, n)$
        **end if**
8:      **end if**
9:    **end for**

---

At this point, with $N$ agent identified phases and $N \times M$ boundary scores calculated, we identify the "most-likely" ground truth label for each agent boundary using an iterative process described in Algorithm S1. This procedure results in each of the $N$ agent identified phases having a ground truth label and boundary score.

## E. Description of *in silico* Active Learning

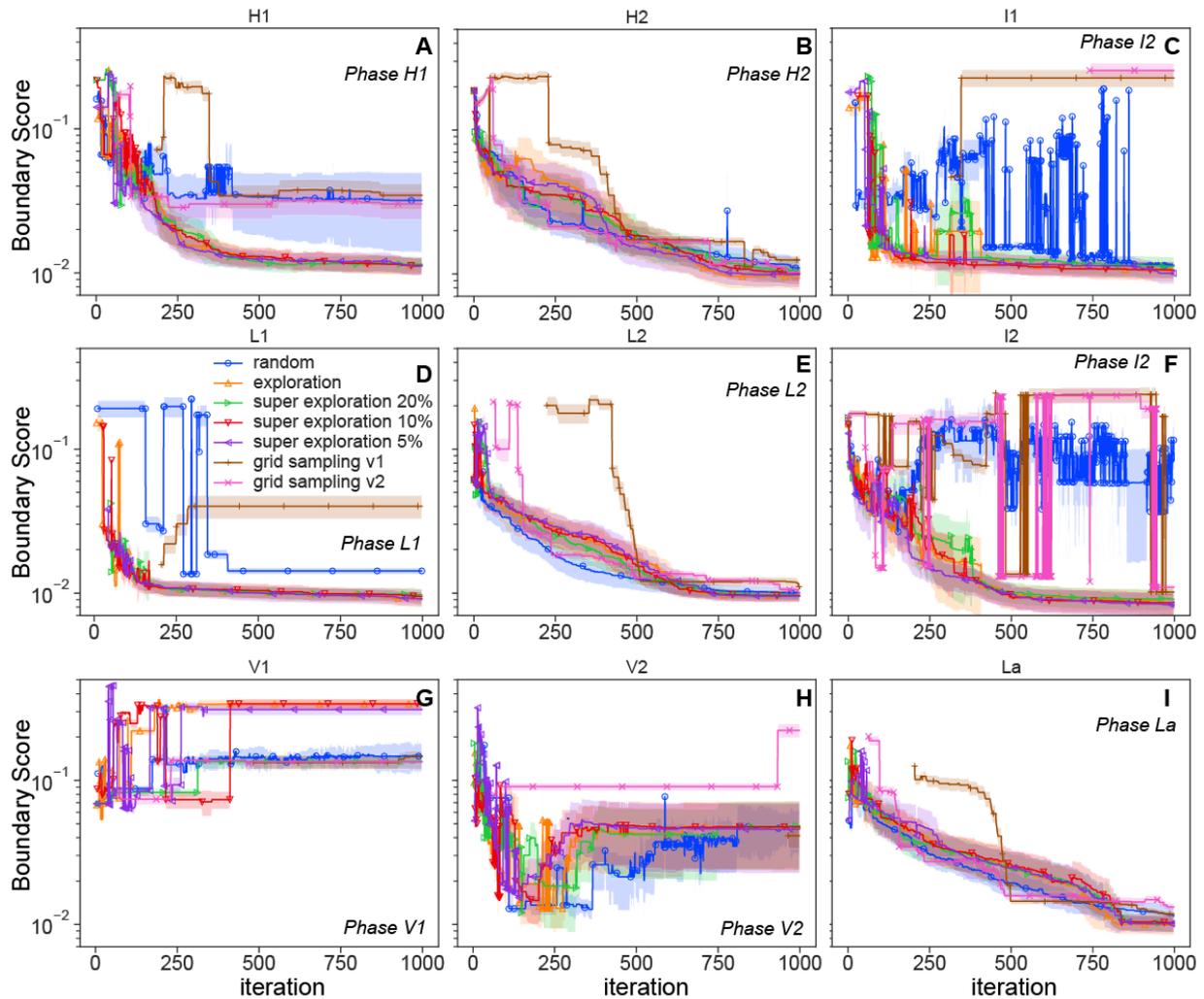

*Figure S11: Performance of the AFL agent in full in-silico active learning runs with different acquisition functions (legend in part D) for each phase in Figure 2B. The boundary score is shown as a function of iteration (number of measurements) for each ground truth phase (see subplot titles). The lines and shaded regions represent the mean and standard deviation of seven independent active learning runs from different initial conditions. Where data is not shown for a specific agent and phase, that agent did not identify that phase at that step in any of our virtual campaigns.*

Here we briefly describe the acquisition functions (AFs) used in these *in silico* tests:

The "random" AF simply chooses a random point from the selection grid using a uniform distribution, ignoring any input from the labeling or extrapolation step. The "exploration" AF works as described in Section I by randomly selecting a point from the top 5 % of the variance calculated from the VGP. The "exploration *n %*" AFs work similarly, except for $n$ % of the steps, the agent randomly chooses a gridpoint without regard to the VGP uncertainty.

For the grid sampling trials, our goal was to mimic how a simple grid scan might be run in an experimental setting. For "grid sampling v1", first 1000 points are selected from the acquisition function grid. From this grid, the agent sequentially measures points at 0 % copolymer, moving from 100 % to 0 % oil starting. Next, the copolymer content is increased by 5 % and the oil % scan is repeated from 100 % to 5 %. Once 100% copolymer is measured, the agent returns to 2.5 % copolymer, and restarts the oil % scan with steps of 5 % copolymer, skipping over any previously measured rows in copolymer %.

The "grid sampling v2" AF uses similar scans from 100 % to 0 % oil but with varying step sizes in copolymer %. In the first pass, 0 %, 50 %, and 100 % copolymer is measured sequentially with 100 % to 0 % oil scans. Then 20 %, 40 %, 60 %, 80 % copolymer is measured, followed by 10 %, 30 %, 70 %, 90 %. Finally, the stepwise increase in copolymer is done for 5 % steps starting first at 5 % and then 2.5 %, skipping any previously measured rows in copolymer %.